\newif\iftodo       
\todofalse
\newif\iftodoshort  
\todoshortfalse
\documentclass[10pt,conference,final,a4paper]{IEEEtran}
\usepackage{amsmath}
\usepackage{amssymb}
\usepackage{theorem}
\usepackage{enumerate}
\usepackage[dvips]{graphicx}
\usepackage{xspace}
\usepackage{nicefrac}
\usepackage[hyphens]{url}
\usepackage{cite}                        
\usepackage{algorithm}
\usepackage{algorithmic}  
\usepackage{dcolumn}
\newcommand{\myline}{\hfill\\} 
\newcommand{\emptypage}%
{
  \newpage
  \vspace*{10cm}
  \pagebreak
}
\newcommand{\textin}{\hangindent0.5cm\hangafter1\myline} 
 
\iftodo
  \usepackage[german,english]{babel}   
  \usepackage{color}
\fi
\theoremstyle{plain}
\theoremheaderfont{\bf}
{\theorembodyfont{\it}
  \newtheorem{theorem}{Theorem}[section]

}
{\theorembodyfont{\rm}
  \newtheorem{definition}[theorem]{Definition}
}
{\theorembodyfont{\rm}

}
\newcommand{\mathcomma}{\:,\:\:}
\newcommand{\mapset}[2]{#1\longrightarrow #2}
\newcommand{\mapto}[2]{#1\longmapsto #2}
\usepackage{dsfont}
   
\newcommand{\RR}{\ensuremath{\mathds{R}}}

\newcommand{\mdef}{:=}

\newcommand{\oneover}[1]{\frac{1}{#1}}

\newcommand{\norm}[1]{\lVert#1\rVert}

\newcommand{\abs}[1]{\lvert#1\rvert}

\newcommand{\Unity}{\ensuremath{\mathbf{1}}\xspace}
\newcommand{\Zero}{\ensuremath{\mathbf{0}}\xspace}

\newcommand{\algassign}{\leftarrow}

\title{Arbitrary Rate Permutation Modulation for the Gaussian Channel}
\author{
\authorblockN{Oliver Henkel}
\authorblockA{%
  Fraunhofer German-Sino Lab for Mobile Communications -- MCI\\ 
  Einsteinufer 37, 10587 Berlin, Germany\\
  \url{henkel@hhi.fraunhofer.de}%
}
}
\begin{document}
\maketitle
\begin{abstract}
In this paper non-group permutation modulated sequences for the Gaussian
channel are considered. Without the restriction to group codes rather than
subsets of group codes, arbitrary rates are
achievable. The code construction utilizes the known optimal group
constellations to ensure at least the same performance but exploit the Gray code
ordering structure of multiset permutations as a selection criterion at the
decoder. The decoder achieves near maximum likelihood performance at low
computational cost and low additional memory requirements at the receiver. 
\end{abstract}
%
\IEEEpeerreviewmaketitle
%
%
%
%
%
%
%
%
%
%
%
%
%
%
\section{Introduction}
The history of permutation modulation goes back more than forty years
where it was first introduced by Slepian in
\cite{slepian-j_procieee65} and more generally in the framework
of group codes in \cite{slepian-j_bell-labs68}. The appropriate definitions
are
\begin{definition}\textin
   Let $x$ denote an unit vector in $\RR^n$ and $G$ a
   finite group (order $M=\abs{G}$) of orthogonal $n$-by-$n$ matrices (or a
   finite orthogonal representation of an abstract group).
   \begin{enumerate}
   \item The pair $(G,x)$ determines the $(M,n)$ \emph{group code}
      with \emph{initial vector} $x$ defined as
      $Gx=\{gx \,|\, g\in G\}$
   \item If $G$ is a permutation group of degree $n$ then the corresponding
      group code is called \emph{permutation code}
   \item If $G$ is the full symmetric group $S_n$ of degree $n$ then the
      corresponding permutation code is called \emph{permutation modulation}
   \end{enumerate}
\end{definition}
Since $G$ consists of orthogonal matrices, $\norm{gx}=\norm{x}=1$ for all
$g\in G$, each codeword can be identified with a point on a $(n-1)$
dimensional sphere. Thus permutation modulation generates spherical
(or equal energy) codes. Usually there are two variants of permutation
modulation in the literature, but here we are concerned with variant I only
as defined above\footnote{Variant II defined in \cite{slepian-j_procieee65}
can be obtained from variant I by applying all possible sign changes in the
components of each $gx$, $g\in G$, whereas now the components of the
initial vector $x$ are assumed to be non-negative}.  
It is explicitly specified as follows \cite{slepian-j_procieee65}:
\begin{equation}
   \label{e.initialx}
   x \mdef (\mu_1^{(m_1)},\dots,\mu_k^{(m_k)})
\end{equation}
where 
$\mu_1<\dots<\mu_k$, 
$\mu^{(i)}$ denotes $i$ repetitions of the value $\mu$, 
and $n=m_1+\dots+m_k$. The code is given by the set of all distinct
permutations of the components of the initial vector $x$. Thus the size of
the code is given as
\begin{equation}
   \label{e.M}
   M \mdef \frac{n!}{m_1!\cdots m_k!} = \frac{\abs{m}!}{m!}
\end{equation}
where the notation on the right hand side has been borrowed from the multi
index notation applied to the vector $m=(m_1,\dots,m_k)$, whereas 
$\abs{m}\mdef m_1+\dots+m_k$ and 
$m!\mdef{m_1!\dots m_k!}$. In this context $x$ can be interpreted as a
multiset, i.e. the set of its components with repetitions and the
corresponding permutations are called multiset permutations. 
The repetitions represented by the vector $m$ have been 
introduced to bound the cardinality $M$ of distinct permutations away from
$n!$, and therefore to support a variety of code rates
\begin{equation}
   \label{e.rate}
  R = \oneover{n}\log_2(M) 
\end{equation}
for a fixed sequence length $n$. 

In this setting two prominent questions have been asked
\begin{enumerate}[Q1)]
\item  Given a group $G$, how to design the initial vector $x$, such that
   the resulting group code has maximum minimal distance? (Here it is
   assumed that each codeword has equal probability, such that the distance
   distribution characterizes the maximum likelihood detection)
\item How to decode efficiently? (Note that the size $M$ in \eqref{e.M} is still
   large in most cases, such that maximum likelihood decoding is not practical)
\end{enumerate}
The first question is known as the \emph{initial vector problem} (IVP) and has
been addressed in \cite{slepian-j_procieee65} through numerical
search only. The further history of the IVP covers in particular the
following items: In
\cite{blake-j_siam72} it has been 
shown that for so-called full homogeneous components of group
representations the IVP can be solved. In the more explicit setting of
$G=S_n$ and prescribed $x$ of the form \eqref{e.initialx} with the vector
$m$ given in advance an optimal solution (i.e. determining the vector
$\mu=(\mu_1,\dots,\mu_k)$) has been obtained in \cite{biglieri.elia-j_it76}.
An algorithm for the general case (arbitrary group $G$) based on
mathematical programming has been introduced in \cite{karlof-j_it89} and
refined in \cite{karlof.chang-j_it97} by means of generalized geometric
programming\footnote{This method is described in detail in
  \cite{chang.karlof-j_comp-ops-res94}, where the authors also remarked, that
  the algorithm does not guarantee to find the global optimum.}.
An explicit analytic solution of the IVP for permutation
modulation ($G=S_n$) has been presented in 
a hardly recognized paper \cite{ingemarsson-j_it90}: Given $(n,k)$ the
optimal initial vector (\ref{e.initialx}) is determined such that the
signal energy is minimized (in particular the vector $m$ is not required to
be given in advance as in \cite{biglieri.elia-j_it76}). Due to the
Lagrangian approach this method finds local minima, but parametrizes all
possible solutions.

For the second question there has been developed an optimal
solution in \cite{slepian-j_procieee65} for the case of permutation
modulation: The receiver replaces the first 
$m_1$ smallest components of the received vector by $\mu_1$, the $m_2$
secondary smallest components by $\mu_2$, and so on. For later reference
let us denote this procedure as the function {\sc SlepianDetect}, which takes a
sequence of length $n$ as input and outputs the sequence just constructed. 
For an arbitrary group $G$ a decoding algorithm has been developed in
\cite{karlof-j_it93}, which refines the function {\sc SlepianDetect} by
performing an iterative search afterwards. Moreover, in
\cite{nordio.viterbo-j_ict03}  
ranking/unranking (resp. demapper/mapper) algorithms for multiset
permutations with respect to lexicographical order are presented, which
allow to convert information bit sequences to codewords and vice versa.

{\bf Contribution:} 
In this paper the following achievements with respect to permutation
modulation (variant I) corresponding to the questions Q1 and Q2 are
presented: 
\begin{enumerate}[{A}1)]
\item 
   The solutions \cite{karlof-j_it89, ingemarsson-j_it90} to the initial
   vector problem provide only quantized rates corresponding to the
   cardinalities of subgroups of the permutation group $S_n$ (or the choice
   of the initial vector respectively), i.e. $M$ is
   exclusively given by (\ref{e.M}). Since $M$ is usually still very large,
   only high rates according to \eqref{e.rate} can be achieved in this way. 
   In this paper a permutation 
   modulation construction scheme allowing \emph{arbitrary rates} while
   maintaining large minimal distances is presented. In particular codes
   with not too large rates in possibly high dimensions $n$ will be
   constructed. These new codes do not possess any group (orbit)
   structure any longer, but can be easily constructed.
\item
   A low complex suboptimal decoder for these codes is presented. This
   decoder, though obtained in a completely different manner, shares some
   properties of the algorithm in \cite{karlof-j_it93} and also with
   \cite{goh.dav.2}. One part of the algorithm is the adaption of the
   mapping/demapping functions in \cite{nordio.viterbo-j_ict03}.
\end{enumerate}
As a further motivation for the particular code constructions presented here serves
a certain application of space-time code design described in
\cite{hen-globecom06}, where space-time block codes are transformed into
spherical codes. 
In this situation a target (space-time code) rate has been specified
together with a (possibly large) space-time code block length $T$. This
code is then transformed into a spherical code with 
sequence length $n>T$, thus the spherical code rate is scaled by a
factor of $\nicefrac{T}{n}$ and one ends up with a quite small rate in a high
dimensional sphere which does not necessary fulfill the requirement (\ref{e.rate}).
\section{Rate adapted code construction}
\label{s.code-construction}
The construction exploits the results obtained in \cite{ingemarsson-j_it90}
about the structure of the optimal initial vector
(\ref{e.initialx})\footnote{Note that the solution is based on some
  Lagrangian optimization technique with discretized constraints. 
  Although not mentioned in \cite{ingemarsson-j_it90}, the proof given 
  exploits the convexity of the Lagrangian functional $f$ to adapt the
  method to the discretized case}: 
For prescribed $k$ and the minimal distance held fixed the optimal initial
vector is a minimizer of the energy functional
\begin{equation}
   \label{e.energyx}
   E(m,\mu) = \sum_{i=1}^k m_i \mu_i^2
\end{equation}
The solution is \cite{ingemarsson-j_it90}
\begin{equation}
   \label{e.sol-energymin}
   \begin{gathered}
      \mu_i = -\frac{k-1}{2}+(i-1)\\
      m_i = \lfloor e^{-(\mu_i^2+\eta)/\lambda} \rceil
   \end{gathered}
   \quad\mathcomma
   i=1..k
\end{equation}
where $\lfloor \cdot \rceil$ denotes rounding to the nearest integer and
$\lambda>0$, $\eta<0$ are some kind of Lagrangian parameters which
parameterize the space of solutions unless $(n,R)$ is fixed according to
(\ref{e.rate}). The corresponding permutation modulation is $S_n\hat x$, where
$\hat x \mdef x/\norm{x}$.

Let us now fix some target rate $R$ and a corresponding code size 
$N=\lceil 2^{nR}\rceil$ ($\lceil\cdot\rceil$, $\lfloor\cdot\rfloor$ denote
the common rounding functions to the next greater resp. smaller integer
number). Then taking a solution 
\eqref{e.sol-energymin} with rate greater than $R$ and corresponding size
$M$ subject to \eqref{e.M} will be the
starting configuration for the next step. This step utilizes ranking algorithms for
multiset permutations. The idea is to find an ordering of multiset
permutations which reflects the Euclidean distance relations between the
corresponding code points. This is roughly achieved by some Gray code
ordering. An algorithm which lists all $M$ multiset permutations with respect
to this ordering can be obtained online from \cite{cos-ruskey.sawada}. 
Having this list at hand, select $N$ out of the total of $M$ elements of
the list as equidistant as possible. To this end 
define $N_0\in \{0,1,\dots,N\}$ to be the maximum number which satisfies
$\lceil M/N \rceil (N-N_0-1)+\lfloor M/N \rfloor N_0 \le M-1$ and set
$n_0 := \lceil M/N \rceil (N-N_0-1)$. With this settings pick the first $N-N_0$ 
elements in the list equidistantly spaced by $\lceil M/N \rceil$ and the
remaining $N_0$ ones starting at position $n_0$ equidistantly spaced by
$\lfloor M/N \rfloor$. The result is an injective mapping
\begin{equation} 
   \label{e.mp-gray-encode}
   \text{{\sc mpGrayEncode}}:\mapset{\{0,\dots,N-1\}}{\{0,\dots,M-1\}}
\end{equation}
which parameterizes $N$ out of the $M$ multiset permutations with largest
possible Gray ordering separation. Let us denote
the set of $N$ selected multiset permutations by $\mathcal{MP}$ and the
corresponding $(N,n)$ code by $\mathcal{C}\subset S_n\hat x$. 
Clearly the minimum distance of $\mathcal{C}$ 
is at least as large as the minimum distance of $S_n\hat x$ and due to the
correspondence between Gray ordering of permutations and Euclidean
distances of codewords, the simple parameterization
\eqref{e.mp-gray-encode} seems promising in order to achieve a large
minimal distance for $\mathcal{C}$. 

To allow for a low complex decoding at the receiver we need a translation
table, which translates lexicographic ranks to Gray code ranks. The
need arises because although we have the algorithm \cite{cos-ruskey.sawada}
which lists all multiset permutations in Gray code order, there is no
corresponding ranking function available (at least to the knowledge of the
author). A rank function for (multiset) permutations is a function, which
assigns to each (multiset) permutation a unique number in the range
$0,\dots n!(M)$ and establishes an ordering of (multiset) permutations. 
The inverse mapping is called an unrank function. 
For ordinary permutations there exist rank and unrank functions with
respect to different ordering criteria, including lexicographic and Gray
code ordering. For multiset permutations lexicographic rank and unrank
functions have been presented in 
\cite[function Demapping/Mapping]{nordio.viterbo-j_ict03}\footnote{Note
  that different from the presentation in \cite{nordio.viterbo-j_ict03} all
  loops and sum boundaries have to be decreased by one for an
  implementation in C, 
  except for the upper sum boundary $l$ in function Mapping. Moreover, j
  has to be chosen as the largest $l$ with the property $s_l\ge0$}
with average complexity proportional to $\nicefrac{nk}{2}$.
Let us denote the lexicographic rank function by {\sc LexRank}. It establishes
the required translation table 
\begin{equation}
   \label{e.lex2gray}
   \text{{\sc Lex2Gray}}:\mapset{\{0,\dots,M-1\}}{\{0,\dots,M-1\}}
\end{equation}
as the inverse mapping of $\mapto{i}{\text{{\sc LexRank}$(\pi(i))$}}$ when 
$\pi(i)$ is the multiset permutation corresponding to $i\in\{0,\dots,M-1\}$ with
respect to the given Gray ordered list and {\sc LexRank}$(i)$ the 
lexicographic rank. This mapping
can be stored efficiently as a list of $M$ integer values at the receiver.
\section{Fast near ML decoding}
The transmission of data proceeds as follows. $nR$ information
bits are mapped to one of the $N$ codewords. Let us assume message $i$ is
assigned to codeword $x(i)\in\mathcal{C}$ given by multiset permutation
no. $i$ in $\mathcal{MP}$ applied to the initial vector $x$. $x(i)$ will be
transmitted through the Gaussian channel, thus the receiver gets
\begin{equation}
   \label{e.transmission-eq}
   y = \sqrt{\rho n} x(i) + w
\end{equation}
where $w\sim\mathcal{N}(\Zero,\Unity)$ denotes a white Gaussian noise
vector and $\rho$ the SNR at the receiver (since $x(i)$ has unit
norm). Algorithm \ref{alg.decoding} presents the decoding procedure in
pseudo code. 
 \begin{algorithm}
 \caption{{\sc Decode}($y$)}
 \label{alg.decoding}
  \begin{algorithmic}[1]
  \STATE MLcandidates $\algassign \emptyset$
  \FOR{$j=0$; $j<2^{k-1}$; $j$++}
  \STATE $y(j) \algassign$ {\sc CreateVariant}$(j)$
  \STATE $z(j) \algassign$ {\sc SlepianDetect}$(y(j))$
  \STATE MLcandidates $\algassign$\\ \quad MLcandidates $\cup$ \{{\sc NewCandidate}$(z(j))$\} 
  \ENDFOR
  \STATE $\hat i \algassign$ MLDecode\{MLcandidates\}
  \RETURN Message no. $\hat i$
  \end{algorithmic}
 \end{algorithm}
The algorithm performs maximum likelihood (ML) decoding with substantially
reduced number of candidate codewords. The careful selection of candidates
is the main achievement of the algorithm. It utilizes and refines the
detection method {\sc SlepianDetect} \cite{slepian-j_procieee65} described already
in the introduction by 
creating appropriate variants $y(i)$ of the received sequence $y$. 

Let us
go into some detail of Algorithm \ref{alg.decoding} now and ignore the
function {\sc CreateVariant} for the moment. Then the loop in line 2 becomes
trivial, $i=0$, $y(0)=y$.
Recall that the function {\sc SlepianDetect} would be equivalent to
ML decoding if we had taken the full codebook $S_n \hat x$ for
transmission. Since our code is a subset, {\sc SlepianDetect} might fail, but the
basic idea of Algorithm \ref{alg.decoding} is that due to the Gray like
ordering of the multiset permutations the obtained candidate is some kind
of neighbor (with respect to Euclidean distance) of the transmitted
sequence (compare the discussion in section \ref{s.code-construction}).
This fact is implicitly contained in the
definition of the function {\sc NewCandidate}, see Algorithm
\ref{alg.new-candidate}:
 \begin{algorithm}
 \caption{{\sc NewCandidate}($z$)}
 \label{alg.new-candidate}
  \begin{algorithmic}[1]
  \STATE $lr \algassign$ {\sc LexRank}$(z)$
  \STATE $gr \algassign$ {\sc Lex2Gray}$(lr)$
  \RETURN $\lfloor \widehat{\text{{\sc mpGrayEncode}}}^{-1}(gr) \rceil$
  \end{algorithmic}
 \end{algorithm}
It takes the output of {\sc SlepianDetect}, calculates its lexicographic rank,
transforms it with the help of {\sc Lex2Gray} (\ref{e.lex2gray}) and estimates 
from that number the codeword number by taking the inverse of
(\ref{e.mp-gray-encode}), where its domain has been enlarged to the reals
(which is denoted by $\hat\cdot$ here in Algorithm
\ref{alg.new-candidate}\footnote{In the spirit of \cite{goh.dav.2}
  the rounding in the last step in Algorithm \ref{alg.new-candidate}
  corresponds to the decision regions provided by an appropriate lookup
  table. However, the use of the Gray code ordering provides fast access to
  the location of $z$ in the lookup table}).

Unfortunately the neighborhood assumption is not correct in general since
transmission errors may occur everywhere in the sequence and this is where
the function {\sc CreateVariant} in step 3 of Algorithm \ref{alg.decoding} enters
the stage. Nevertheless Algorithm
\ref{alg.new-candidate} provides codeword candidate numbers from scratch without
going through the list of all codewords in $\mathcal{C}$, thus its
computational complexity is very low. The final detection step 7 denotes ML
detection with respect to the codewords listed in the set
MLcandidates. Since the cardinality of this set is at most $2^{k-1}$ the
final ML detection has low complexity also.
 
At last let us consider the function {\sc CreateVariant}s. The idea is, that at
least for not too low SNR values, errors occur only in a few places. Recall
that since each ordered $m_i$ received values will be equated to $\mu_i$ in
{\sc SlepianDetect}, thus an error occurs only, if the smallest or largest of
them is perturbed so badly, that {\sc SlepianDetect} assigns a wrong value
$\mu_j$, $i\neq j$ to it. So this defect interchanges the components $m_i$
and $m_i+1$ in the sorted sequence. The function {\sc CreateVariant} loops over
all $k-1$ such places and interchanges the corresponding components,
compare Algorithm \ref{alg.create-variant}.
This strategy obviously approaches ML performance when the SNR grows.
\begin{algorithm}
   \caption{{\sc CreateVariant}($j$)}
   \label{alg.create-variant}
   \begin{algorithmic}[1]
      \STATE create binary representation $j=\sum_{l=1}^{k} b_l 2^{l-1}$
      \STATE $\forall_{l\,|\,b_l=1}$: interchange component no.
             $m_l$ and $m_l+1$ in the sorted version of $y$
   \end{algorithmic}
\end{algorithm}
\section{Simulations and discussion}
%
%
%
%
%
%
%
%
%
%
%
%
%
%
%
%
%
%
%
%
%
\addtolength{\textheight}{1cm}   
The following codes have been constructed:
\begin{enumerate}
\item A code with sequence length $n=25$ with $N=323$ codewords (supporting
   a target rate of $\nicefrac{1}{3}$) out of $M=600$ multiset permutations
   corresponding to $m=(1,23,1)$ and
   $\mu=(-\oneover{\sqrt{2}},0,\oneover{\sqrt{2}})$. 
\item This code has sequence length $n=50$, $N=1024$ (supporting the target
   rate $\nicefrac{1}{5}$) out of $M=2450$ multiset permutations
   corresponding to $m=(1,48,1)$ and
   $\mu=(-\oneover{\sqrt{2}},0,\oneover{\sqrt{2}})$.
\item The last code has sequence length $n=100$, $N=1024$ (supporting the target
   rate $\nicefrac{1}{10}$) out of $M=9900$ multiset permutations
   corresponding to $m=(1,98,1)$ and
   $\mu=(-\oneover{\sqrt{2}},0,\oneover{\sqrt{2}})$.
\end{enumerate}

\begin{figure}[htb]
   \includegraphics[scale=0.3]{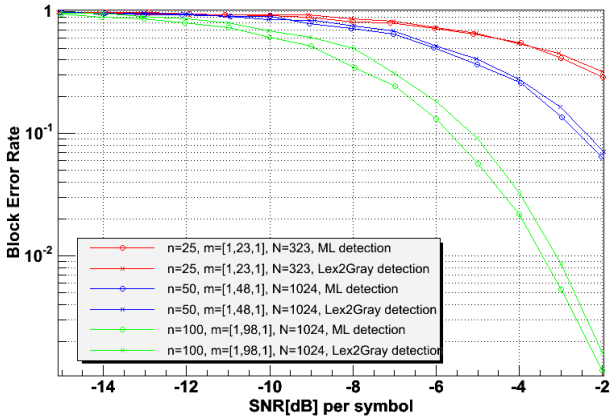}
   \caption{Performance of codes 1-3 with ML detection and detection
     corresponding to Algorithm \ref{alg.decoding}\label{fig.ml-vs-l2g}}
\end{figure}

Their performance is shown in Fig. \ref{fig.ml-vs-l2g} for the low SNR regime.

\begin{figure}[htb]
   \includegraphics[scale=0.3]{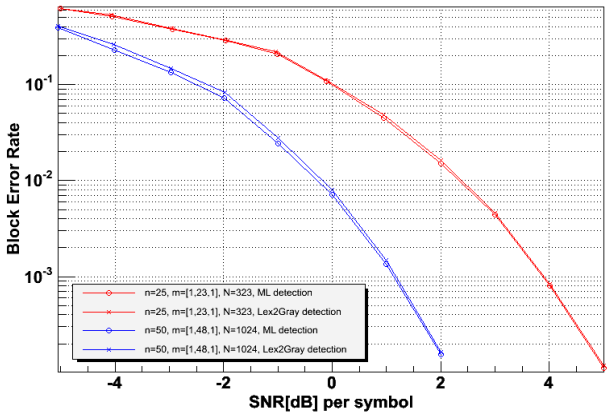}
   \caption{Performance of codes 1 and 2 in the mid SNR regime with ML
     detection and detection 
     corresponding to Algorithm \ref{alg.decoding}\label{fig.mid-snr}}
\end{figure}

The mid SNR regime for the codes 1 and 2 is shown in
Fig. \ref{fig.mid-snr}. The last simulations approve the expectation that
the proposed decoding 
algorithm approaches ML performance when the SNR grows. In fact the
performance curves match quite well. 

Observe that ML decoding amounts for code 3 1024 comparisons of 100
dimensional vectors, while Algorithm \ref{alg.decoding} singles out only 4
candidates for ML decoding which are obtained by simple element
operations. The complexity of Algorithm \ref{alg.new-candidate} is 
approximately $nk/2=150$, thus the complexity of the whole decoding
procedure (Algorithm \ref{alg.decoding}) scales with 
$n\log(n)$ (due to the sorting operations involved) and is independent of
$N$ in contrast to ML decoding which 
scales with $Nn$, whereas usually $N\gg n$ holds. As an illustration of
computational complexity, table 
\ref{tab.cputimes} lists the CPU time for the whole simulation process of
codes 1-3 in the SNR range from -15dB to -2dB. Each simulation has been
performed on the same machine applying Algorithm \ref{alg.decoding} and ML
decoding respectively. The table reveals 97.5 up to 99 percent saving.
\newcolumntype{.}{D{.}{.}{4}}
\begin{table}[htb]
   \begin{center}
   \begin{tabular}{|c|.|.|.|}
      \hline
                                   & \text{code 1} & \text{code 2} & \text{code 3}  \\
      \hline
      Algorithm \ref{alg.decoding} & .38    & 2.22   & 128.54  \\
      \hline
      ML decoding                  & 14.73  & 280.46 & 20346.7 \\
      \hline
      ratio                        & .0258  & .0079  & .0063   \\
      \hline
   \end{tabular}
   \caption{Comparison of CPU simulation times with and without near ML
     decoding \label{tab.cputimes}} 
   \end{center}
\end{table}

In summary, the encoding method described in this paper have been shown to 
support arbitrary rates with minimum distance lower bounded by the optimal IVP
in the sense of \cite{ingemarsson-j_it90} with respect to the underlying
permutation modulation. Complementary the presented decoding algorithm
achieves near ML performance with substantial savings of computation load
at the receiver and only small additional memory requirements.
By the way a practical implementation for determining the
Gray code rank of multiset permutations has been achieved with the help of
a simple translation table, which provides fast access. 
Note also, that encoding and decoding both are simple in structure. 
%
%
%
%
%
%
%
%
%
%
%
\bibliography{%
/home/henkel/texstuff/bib/ieee/IEEEabrv,%
/home/henkel/texstuff/bib/refmci,%
/home/henkel/texstuff/bib/myrefs} 
\end{document}